\documentclass[a4paper,11pt]{article}
\usepackage[textwidth=16cm]{geometry}
\usepackage[font=footnotesize,labelfont=bf]{caption}

\usepackage{amsmath, amssymb, graphicx}
\usepackage{color, xspace}
\usepackage{algpseudocode, algorithm, algorithmicx}
\usepackage{soul}
\usepackage{authblk}
\usepackage{placeins}
\usepackage{dblfloatfix}

\newcommand{\person}[1]{\textsc{#1}}

\newcommand{\nn}{\nonumber}
\newcommand{\la}{\langle}
\newcommand{\ra}{\rangle}
\newcommand{\myvec}[1]{{\bf#1}}
\newcommand{\ddt}{\frac{\partial}{\partial t}}
\newcommand{\micron}{\mu{\rm m}}

\newcommand{\Smoldyn}{\textit{Smoldyn}\xspace}
\newcommand{\ChemCell}{\textit{ChemCell}\xspace}
\newcommand{\MCell}{\textit{MCell}\xspace}
\newcommand{\GridCell}{\textit{GridCell}\xspace}
\newcommand{\ReaDDy}{\textit{ReaDDy}\xspace}
\newcommand{\GFRD}{GFRD\xspace}
\newcommand{\eGFRD}{eGFRD\xspace}
\newcommand{\pGFRD}{pGFRD\xspace}
\newcommand{\eCell}{\textit{eCell}\xspace}
\newcommand{\FPKMC}{FPKMC\xspace}
\newcommand{\SVTA}{SVTA\xspace}
\newcommand{\Spatiocyte}{\textit{Spatiocyte}\xspace}
\newcommand{\pSpatiocyte}{\textit{pSpatiocyte}\xspace}
\newcommand{\AND}{\textbf{and} }
\newcommand{\NOT}{\textbf{not} }

\newcommand{\ToAppearNote}{to appear in:  B. Munsky, W.S. Hlavacek, and L. Tsimring (editors), {\it ``Quantitative Biology: Theory, Computational Methods and Examples of Models''}, MIT Press, Cambridge, MA, U. S. A. (expected in early 2018)}

\newcommand{\FigDir}{.}

\begin{document}
\title{Spatial-Stochastic Simulation\\of Reaction-Diffusion Systems\footnote{\ToAppearNote}}

\author[1]{Thomas R. Sokolowski\thanks{e-mail: \texttt{tsokolowski@ist.ac.at}}}
\author[2]{Pieter Rein ten Wolde}
\affil[1]{\footnotesize IST Austria, Am Campus 1, A-3400 Klosterneuburg, Austria}
\affil[2]{\footnotesize FOM Institute AMOLF, Science Park 104, NL-1098 XG Amsterdam, The Netherlands}

\date{\today\vspace{-1em}}

\maketitle

\abstract{
In biological systems, biochemical networks play a crucial role,
implementing a broad range of vital functions from regulation and communication
to resource transport and shape alteration.
While biochemical networks naturally occur at low copy numbers and in a spatial setting,
this fact often is ignored and well-stirred conditions are assumed for simplicity.
Yet, it is now increasingly becoming clear that even microscopic spatial inhomogeneities can profoundly influence
reaction mechanisms and equilibria, oftentimes leading to apparent differences on the macroscopic level.
Since experimental observations of spatial effects on the single-particle scale are extremely challenging under {\it in vivo} conditions,
theoretical modeling of biochemical reactions on the single-particle level is an important tool 
for understanding spatial effects in biochemical systems.
While the combined requirement of incorporating space and stochasticity quickly limits the tractability of purely analytical models,
spatial-stochastic simulations can capture a wide range of biochemical processes with the necessary minimal levels of detail and complexity.
In this chapter we discuss different simulation techniques for spatial-stochastic modeling of reaction-diffusion systems, 
and explain important working steps required to make them biochemically accurate and efficient.
We illustrate non-negligible accuracy issues arising even in the most simple approaches to biochemical simulation,
and present methods to deal with them.
In the first part of the chapter we explain how {\it Brownian Dynamics}, a widely used particle-based diffusion simulation technique 
with a fixed propagation time, can be adapted to simulate chemical reactions as well, 
and portray a range of simulation schemes that elaborate on this idea.
In the second part, we introduce event-driven spatial-stochastic simulation methods, in which simulation updates are performed asynchronously 
with situation-dependent, varying time steps;
here we particularly focus on {\it eGFRD}, a computationally efficient particle-based algorithm that makes use of analytical functions 
to accurately sample interparticle reactions and diffusive movements with large jumps in time and space.
We end by briefly presenting recent developments in the field of spatial-stochastic biochemical simulation.
}

\newpage
\tableofcontents

\newpage
\section{Why spatiality matters}
\label{Sec-Sokolowski-Intro}
As a matter of course, all chemical reactions occur in space.
However, in classical chemical theory, where chemical reaction systems are traditionally modeled with deterministic mass-action kinetics,
space only enters the picture as a normalization constant in form of the volume, which implicitly is assumed to be homogeneous and of irrelevant geometry.
This allows operating with macroscopic quantities such as the concentration and macroscopic (phenomenological) reaction rates;
so long as concentrations are high enough such that the average distance between reaction partners is small,
mass-action models will describe the given chemical system accurately enough.
In biological systems, however, this premise cannot be taken for granted.
The typical situation in cells is characterized by a large variety of biochemical species that---more often than not---are present at astonishingly low numbers;
concentrations in the $\mu{\rm M}$ and nM regime are not seldom, and in particular proteins acting as transcription factors can reach copy numbers as low as 10 per cell.
Moreover, cells have evolved different strategies of symmetry breaking and compartmentalization that---on purpose---efficiently localize reactions to cellular substructures
such as the cell membrane, cell organelles, cytoskeletal filaments and scaffold proteins.
A high degree of spatial inhomogeneity and significant deviations from the behavior predicted by mass-action kinetics is the consequence.
Most importantly, in the limit of low copy numbers, when reactant distributions become highly non-uniform and inter-reactant distances large, 
the stochastic transport phenomena that lead to reactant encounters can become more important than particle abundance itself.

To name a few examples for the importance of spatial aspects in biochemical reaction-diffusion systems, partly revealed by the application of the techniques explained in this chapter:
Spatial inhomogeneities can have a strong effect on the behavior of spatially distributed enzymes \cite{ELF:2004SystBiol,LAWSON:2015JRSocInterface},
going as far as provoking the emergence or destruction of ultrasensitivity \cite{VAN_ALBADA:2007PLoSComputBiol,MORELLI:2008BiophysJ,TAKAHASHI:2010PNAS,DUSHEK2011:BiophysJ},
and on (density-dependent) clustering \cite{JILKINE:2011PLoSComputBiol,WEHRENS:2014JChemPhys}.
Macromolecular crowding can shift chemical equilibria \cite{MORELLI:2011BiophysJ,KLANN:2009BiophysJ,KLANN:2011BMCSystBiol} (see \cite{TEN_WOLDE_MUGLER:2014IntRevCellMolBiol} for a review), 
and fast reactant rebindings can significantly enhance the noise in transcription factor and ligand binding \cite{VAN_ZON:2006BiophysJ,MUGLER:2012BiophysJ,KAIZU:2014BiophysJ}.
Facilitated diffusion on one-dimensional submanifolds, such as the DNA or cytoskeletal macropolymers, is capable of facilitating the search for target sites \cite{PAIJMANS:2014PhysRevE,LOVERDO:2008NatPhys,BENICHOU:2010NatChem},
while membrane partitioning can result in marked enhancement of membrane-transduced signals \cite{MUGLER:2013PNAS}.
Most strikingly, spatio-temporal fluctuations at the molecular scale can drastically change the macroscopic behavior on the cellular scale \cite{TAKAHASHI:2010PNAS,FANGE:2006PLoSComputBiol,MUGLER_TEN_WOLDE:2013AdvChemPhys}.
Not least, spatial averaging often substitutes or complements temporal averaging in noisy signal readouts \cite{ERDMANN:2009PRL,SOKOLOWSKI:2012PLoSComputBiol,SOKOLOWSKI:2015PhysRevE}.

As a first step to model spatial inhomogeneity in chemical systems, one is lead to partition the previously uniform reaction volume into small subvolumes, each tracking a local concentration variable. This is the strategy of RDME\footnote{Reaction-Diffusion Master Equation}-based simulation algorithms, that are briefly described in Sec.~\ref{Sec-Sokolowski-RDME} of this chapter, and in more detail in Ch.~22; these approaches, however, still assume well-stirredness locally, and when particle numbers indeed reach as low as 10 this must ultimately break down as well. In that limit, any reasonable modeling approach must account for individual particle positions and the stochastic transport processes that alter them.
While this introduces an inevitably higher degree of complexity, the design of suitable algorithms to simulate such models is aided by the fact that the reactant particles are typically transported by diffusion---a process whose statistics is well understood.

\begin{figure}[t]
  \centering
  \includegraphics[width=\textwidth]{\FigDir/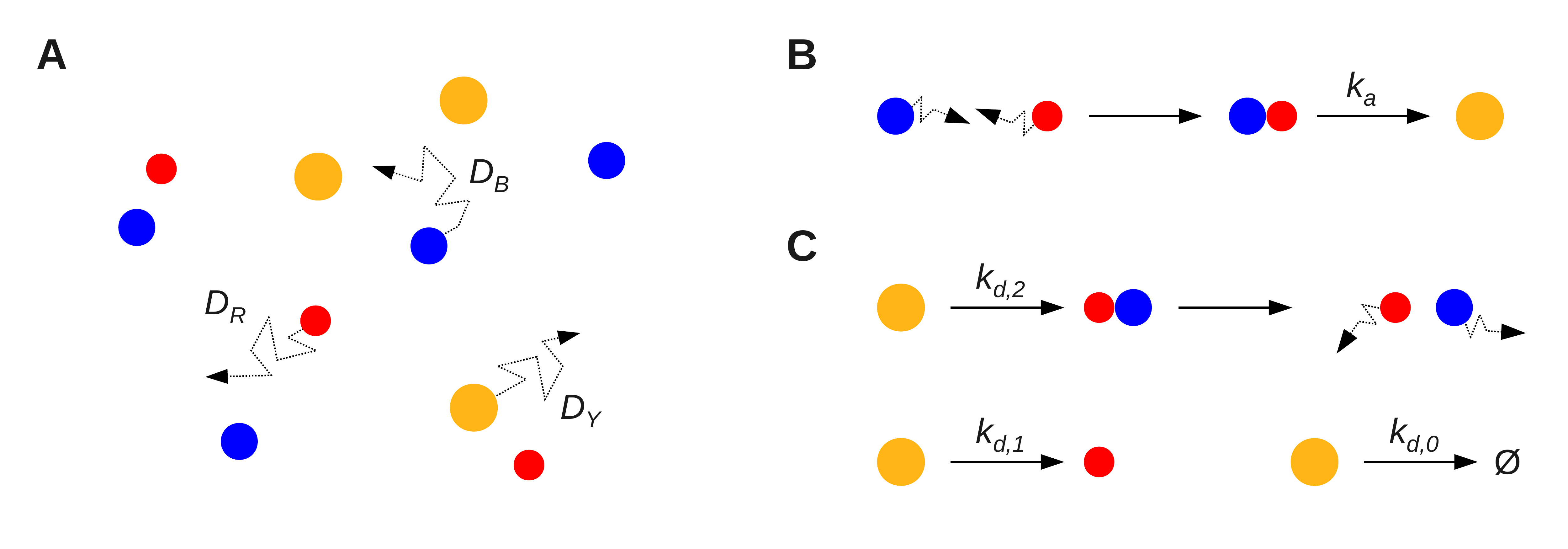}  
\caption{ \label{figStdModel}
  \textbf{Stochastic particle-based model of a reaction-diffusion system.}
	  (A) In the arguably most simple abstraction, particles of different biochemical species (named $B$, $R$ and $Y$) are represented as 3D solid spheres with species-specific radii.
	  We mark different chemical species by different colors; here and throughout we use the term ``species'' in a generic sense, not only referring to the chemical composition of particles, but rather categorizing them according to all properties that affect their reaction-diffusion behavior.
	  The particles can randomly diffuse in space with their species-specific diffusion coefficient.
	  Note that while here we indicate this by discrete random walk paths, we imagine the diffusion process to be continuous.
	  We show here a 2D projection for the standard scenario in which particles freely diffuse and react in 3D space.
	  The model can be straightforwardly extended towards diffusion in lower dimensions and in confined geometries.
	  (B) When particles randomly come close enough to establish a ``contact situation'' they can undergo a bimolecular reaction
	  to form a product with an ``intrinsic'' association rate $k_a$; note that this rate conceptually differs from the ``macroscopic'' rate appearing in mass-action kinetics.
	  (C) Particles can undergo unimolecular reactions, either reacting back to their educts with dissociation rate $k_{d,2}$ (upper reaction),
	  changing their species with rate $k_{d,1}$ (lower left), or being annihilated with decay rate $k_{d,0}$ (lower right).
	  Upon annihilation, a particle is removed from the system.
}
\end{figure}

\pagebreak
But how much more of complexity do we need to invoke to make relevant effects on the particle scale visible?
In Fig.~\ref{figStdModel} we introduce the arguably most simple stochastic reaction-diffusion model retaining particle positions, 
which in the following we will refer to as the ``stochastic
particle-based model'', or simply ``the particle-based model''.
Here, particles are represented by solid spheres whose radii depend on the chemical species; throughout this chapter we will use the term ``species'' in a broad sense, not only to distinguish particles by different chemical composition, but rather by all properties altering their reaction-diffusion behavior, such as different conformations and modification states.
The particles can diffuse in space with a species-specific diffusion constant; here we typically think of 3D-diffusion, but adaptations to lower dimensions are straightforward.
Once two particles randomly approach each other close enough to establish a ``contact situation'', they can undergo a bimolecular reaction and form a product with a forward reaction rate $k_a$;
this ``intrinsic'' reaction rate is conceptually different from the usual ``macroscopic'' rate as known from mass-action kinetics.
In practice---as will become clear later in this chapter---the right choice of $k_a$ not only depends on the species of the involved particles, but also on the precise definition of the ``contact'' in order to correctly reproduce chemical equilibria.
Obviously, correct sampling of the equilibria also requires including the back reactions.
The model therefore also allows for unimolecular reactions which can either result in dissociation of a product particle into its educts,
a species change (representing, e.g., an allosteric transition), or complete decay of the particle in which case it is taken out of the system;
in principle, any particle can have an arbitrary number of such reactions, while all unimolecular rates can be different.
Note that the minimal model introduced here ignores interparticle forces and the internal structure of the particles, which can influence the chemical behavior of the system;
we comment on extended schemes that consider these features in Sec.~\ref{Sec-Sokolowski-ReaDDy} and Sec.~\ref{Sec-Sokolowski-Hybrid}, respectively.

In this chapter, we will introduce the general principles underlying various types of particle-based stochastic algorithms capable of simulating the above model,
and portray toolkits that implement them, while pointing out their different advantages and caveats.
Among the available algorithms one can distinguish two classes grouping conceptually similar schemes: algorithms based on Brownian dynamics (BD), which update particle states with a fixed time step, described in Sec.~\ref{Sec-Sokolowski-BD}, and event-driven algorithms such as \eGFRD, aiming at predicting next-event times for ``interesting'' events (i.e., the reactions) between ``uninteresting'' periods of diffusive motion, introduced in Sec.~\ref{Sec-Sokolowski-EventDriven}.
In Table~\ref{Tab-Algorithms} we briefly compare advantages and disadvantages of the two main algorithm classes and the already mentioned RDME-based algorithms,
in terms of practically relevant criteria.

\begin{table}
 \centering
 \begin{tabular}{|l|c|c|c|}
  \hline
	    & BD schemes & (e)GFRD & RDME-based\\
  \hline
  Accuracy				&&&\\
  ~~- spatial detail			&very good	&very good	&medium \\
  ~~- microscopic reaction kinetics	&good for some	&very good	&poor 	\\
  \hline
  Computational efficiency		&&&\\
  ~~- low concentrations 		&low		&high		&high 	\\
  ~~- high concentrations 		&low		&low		&high 	\\
  \hline
  Overhead				&&&\\
  ~~- mathematics/analytics effort	&medium	to low	&very high	&low	\\
  ~~- implementation effort		&low		&high		&medium	\\
  \hline
 \end{tabular}
\caption{ \label{Tab-Algorithms}
Comparison of (dis)advantages for different classes of spatial-stochastic biochemical simulation schemes, according to relevant criteria.
Note that accurate modeling of bimolecular reactions in BD schemes requires considerable analytical effort; naive BD schemes therefore differ from advanced schemes both in accuracy and overhead (cf. Sec.~\ref{Sec-Sokolowski-BD}).}
\end{table}

\FloatBarrier

\pagebreak
\section{Brownian dynamics simulations with reactions}
\label{Sec-Sokolowski-BD}

The arguably easiest and most intuitive approach to particle-based biochemical simulation are Brownian dynamics (BD) simulations.
Here, the basic concept is straightforward: 
Let us assume that the particles are represented by solid spheres and thus entirely characterized by their species, position and radius,
as in the model of Fig.~\ref{figStdModel}.
In BD schemes the diffusive motion of the particles is approximated by a discrete random walk.
To that purpose, upon each update the particles are moved by small, Gaussian distributed displacements, using a fixed time step $\Delta t$.
Upon particle collision, representative of the ``contact situation'' in the model,
it is checked whether the particles will react within the time step $\Delta t$ or not.
If so, the two colliding particles are replaced by a product particle.
If no reaction occurs, the particles will be displaced again and move apart in the next step.

The small displacements $\Delta\myvec{r}$ are sampled from the
  Gaussian probability density function for the travelled distance of
  a diffusing particle after the time $\Delta t$,
\begin{align}
\label{Eq-Sokolowski-FreeGF}
p(\Delta\myvec{r}, \Delta t) &= \frac{1}{\left(4\pi D \Delta t\right)^{d/2}} e^{-|\Delta\myvec{r}|^2 / (4 D \Delta t)} ~,
\end{align}
where $d$ is the dimension of the diffusive process.
In order to be able to check for particle collisions, one has to ensure that the typical displacement will be smaller than the size of the particles, meaning that $\overline{\Delta\myvec{r}}\equiv\la |\Delta\myvec{r}|^2 \ra^\frac{1}{2} = \sqrt{2dD^*\Delta t} \ll R^*$, where $R^*$ is the largest particle radius, and $D^*$ the highest diffusion coefficient.
From this it follows that $\Delta t \ll {R^*}^2 / (2d D^*)$.
Thus, since $R^*$ typically is of nanometer order and not overly variable accross different particle types, 
high diffusion coefficients limit the choice of the maximal time step.

When particles end up overlapping after an update, in the next step it is checked whether they react.
The probability for the reaction to occur within the time interval $\Delta t$, to a first approximation, is given by
\begin{align}
 p_r(\Delta t) &= k_a \Delta t ~,
\end{align}
where $k_a$ is the intrinsic rate for reaction of the two species at particle contact.
Obviously, a necessary condition for sampling this probability (using the standard method of comparing to a random number $\mathcal{R}\in [0,1]$) is $p_r \leq 1$, 
implying $\Delta t \leq 1/k_a$.
Since this must hold for all reactions involved, the fastest rate $k_a^*$ will also restrict the maximal value of $\Delta t$, in addition to $D^*$ and $R^*$ above.

Unimolecular reactions such as particle decay and internal state changes---assuming that these processes follow Poissonian statistics---can be sampled by comparing uniform random numbers to the probability that the event occurs during the time period $\Delta t$,
\begin{align}
p(\Delta t) = 1 - e^{-k_d \Delta t} ~,
\label{Eq-Sokolowski-Unimolecular}
\end{align}
where $k_d$ is the corresponding unimolecular reaction rate.

In most applications of interest, at least one of the involved rates or diffusion coefficients will be ``fast'', i.e. differ significantly from zero.
Therefore, time steps as low as $\Delta t \sim 10^{-9}s$ are not uncommon, implying large numbers of updates required to produce noticable changes and advances in simulated time.
Particularly in ``sparse'' systems, in which particle distances are large compared to particle radii, such small time steps will mean that most of the computational effort will be spent on sampling the small diffusive displacements. This renders BD simulations very inefficient in such situations, which are not untypical in biology.
Later, in Sec.~\ref{Sec-Sokolowski-EventDriven}, we will introduce \eGFRD, an event-driven simulation algorithm that resolves this drawback of BD elegantly and efficiently.

Another caveat of BD simulations arises from the fact that reaction attempts are always made from contact situations whose definition, to a certain extent, is arbitrary;
consequently, the microscopic reaction rates conditioned on particle contact have to be tweaked such that macroscopic equilibria are correctly reproduced.
When particles are abstractly modeled as perfect spheres, one typically defines the contact situation 
as the one in which the interparticle distance equals the sum of the particle radii (particles touching), in accordance with intuition.
However, in BD simulations in which particle displacements are always discrete, such ``perfect'' contact almost never occurs;
rather, the particles ``at contact'' will practically always have a finite overlap.
This introduces a small error that, strictly speaking, renders BD simulations inexact;
choosing small $\Delta t$ will attenuate, but never completely remove the error.
Even more significantly, as a consequence of this error, detailed balance is broken when decay products in reversible reactions are naively placed at perfect contact in the back reaction; this systematic error will accumulate as these reactions repeat, and can noticeably alter the sampled equilibria.
Further below, in Sec.~\ref{Sec-Sokolowski-RBD}, we present two improved schemes that restore detailed balance in BD simulations.

Notwithstanding their limitations, BD simulations have a striking advantage: thanks to their comparably simplistic simulation algorithm, they are easily implemented and allow for simulations in complex geometries, as soon as local interaction rules have been accurately specified. This also enables a rather straightforward inclusion of force-interactions between particles and other objects, as described in Sec.~\ref{Sec-Sokolowski-ReaDDy}.

Above we explained that naive BD simulation schemes necessarily must trade-off accuracy against efficiency.
We will now discuss in more detail several variations on the basic BD simulation principle, partly designed to overcome the abovementioned shortcomings.

\subsection{Smoldyn, MCell and similar schemes}
\label{Sec-Sokolowski-SmoldynMCell}
The recognition of the nontrivial challenges connected to correct sampling of the reaction events in naive BD schemes, outlined in the previous section, 
has led to various algorithms that propose different solutions for this issue.
Below we discuss two popular ones, \Smoldyn \cite{ANDREWS:2004PhysBiol,ANDREWS:2010PLoSCompBiol} and \MCell \cite{STILES:2000CompNeurosci,KERR:2008SIAMJSciComput}, and briefly mention related approaches.

One framework that can simulate reversible reactions is \Smoldyn
\cite{ANDREWS:2004PhysBiol,ANDREWS:2010PLoSCompBiol}.  The solution put forward here is to replace the reaction/binding radius of two particles by an effective value that is smaller than the sum of the particle radii whenever the reaction rate is not infinite, i.e. whenever reactions are not purely diffusion-limited.
The particles then \textit{always} react at contact, but the contact radius is reduced precisely in such way that the macroscopic reaction rates are matched. 
This approach has the following issues:
First of all, the interaction cross-section is not determined by the size of the particles, but depends in a rather ad-hoc fashion on the overall reaction rate:
for purely diffusion-limited reactions, the effective cross section in the simulations is given by the actual cross section corresponding to the physical size of the particles, 
but for reactions that are in the reaction-limited regime, the effective cross section goes to zero, becoming much smaller than the actual cross section. 
At low concentrations, where particle rebindings can be integrated out \cite{VAN_ZON:2006BiophysJ,KAIZU:2014BiophysJ}, this approach may be feasible.
However, when the dynamics at the molecular scale becomes vital, as {\it e.g.} in protein clusters, in reactions at or near the membrane \cite{MUGLER:2012BiophysJ}, 
or in scenarios with multi-site protein modification, where rapid enzyme-substrate rebindings even can qualitatively change the macroscopic behavior of the system \cite{TAKAHASHI:2010PNAS}, 
the approach of resizing effective particle radii should be used with great care---it is far from obvious that in such cases it can describe the dynamics correctly, even qualitatively. 
Moreover, it has also a practical consequence:
the small effective cross section for reaction-limited reactions demands a correspondingly small propagation time step,
in order to ensure that the typical distance traveled by particles per update is smaller than their effective size;
this lowers the computational efficiency.
The second issue relates to the fact that, in the back reaction, the educt particles cannot be placed back in the exact spatial configuration from which the forward reaction occured,
i.e. at the (artificial) contact distance set by the effective cross section $\sigma_b$, because then the algorithm would force them to re-form the product instantly in the next step.  
This requires placing dissociating particles back at an (equally artificial) ``unbinding radius'' $\sigma_u > \sigma_b$, which violates detailed balance.
Despite its limitations, Smoldyn has been applied to various systems, including the {\it E. Coli} chemotactic sensing network \cite{LIPKOW:2005JBacteriol,LIPKOW:2006PLoSComputBiol,LIPKOW:2008CellMolBioeng},
constrained diffusion in mitosis \cite{BOETTCHER:2012JCellBiol,ZAVALA:2014PLoSComputBiol}, membrane-associated reactions \cite{GERISCH:2013JCellSci,HOFFMANN:2014SoftMatter,MCCABE_PRYOR:2013BiophysJ}, and problems in neuroscience \cite{GRATI:2006JNeurosci,SINGH:2011PLoSComputBiol,KHAN:2011JComputNeurosci,KHAN:2012JComputNeurosci}.

\MCell \cite{STILES:2000CompNeurosci,KERR:2008SIAMJSciComput,STEFAN:2014PLoSComputBiol} originally was designed to simulate surface reactions in neurotransmission, initially on planes, in recent versions also on triangulated surfaces;
it therefore puts a particular emphasis on bulk-surface reactions.
As a specialty, \MCell takes the approach of ``ray-marching'' the trajectories of diffusing particles within the next time step to detect collisions in (potentially complex) subvolumes,
much inspired by ray-tracing techniques from photorealistic computer graphics.
The \MCell rays do not represent the (quickly randomized) ballistic movements of the particle on the molecular scale of water ($\lesssim$ \AA{}), but rather are thought to approximate a particular random diffusion trajectory ``deflected'' by impermeable obstacles on the mesoscopic subcellular scale ($\gtrsim{\rm nm}$).
Using the raytracing approach to ensure microscopic (spatial) reversibility, \MCell puts care into determining the correct distribution of particles after dissociation.
While in \MCell particle interaction radii are not rescaled, the calculation of reaction acceptance probabilities follows a similar spirit as in \Smoldyn:
the acceptance probability for reaction along a ray-marched particle trajectory is determined by analytically computing the expected number of collisions within a cylndrical, interaction-radius dependent volume along the ray, and relating it to the macroscopic reaction rate.
Therefore, \MCell also does not capture the microscopic behavior on the scale of the particle radii in its full richness;
in particular, fast rebindings cannot be sampled with full accuracy, because the probability of rebinding in the next step is computed from considerations 
that do not take into account the distance of the (just dissociated) reactants.
\MCell has found applications mainly in its field of origin (neurobiology) \cite{STILES:1996PNAS,DIAMOND:2001JNeurosci,WACHMAN:2004JNeurosci,COGGAN:2005Science,BEENHAKKER:2010JNeurosci,NADKARNI:2010PLoSComputBiol}, but also in bacterial chemotaxis models \cite{RAPPEL:2008PNAS,RAPPEL:2008PRL}, models of nuclear import \cite{BECK:2007Nature}, and in testing the \person{Berg-Purcell} theory of chemical sensing accuracy \cite{WANG:2007PhysRevE,BERG:1977BiophysJ}.

For completeness, we also mention a number of other BD-based simulation environments: \ChemCell \cite{PLIMPTON:2005JPhysConfSer}, \GridCell \cite{BOULIANNE:2008BMCSystBiol} and \Spatiocyte \cite{ARJUNAN:2009SystSynthBiol}.
\ChemCell is one of the few toolkits that supports both spatial (BD) and non-spatial (\person{Gillespie}-type) simulations since its first release.
The BD simulation algorithm of \ChemCell is very much of the kind as in \Smoldyn and \MCell.
In \GridCell, particles are propagated via random walks on a static voxel grid (the D3Q27 grid).
This significantly facilitates collision detection because the neighborhood of each particle is stored in the lattice and thus known at any update.
On the other hand, operating on the lattice necessitates to adapt the lattice parameters such that the random walk reproduces the biophysical diffusion constant in the macroscopic limit;
likewise, since the microscopic reaction events are simulated in an ad-hoc fashion, reaction acceptance probabilities have to be adapted to match the macroscopic rates.
As a consequence, \GridCell represents the microscopic biochemical mechanisms rather poorly, 
and fast diffusion constants or reaction rates will dictate very small time steps or coarse lattice spacing.
Its advantage thus mainly consists of the easy particle neighborhood management, 
which increases computational efficiency in crowded situations, and facilitates parallelization.
The idea of particle-based reaction-diffusion simulations on a spatial lattice is also implemented by \Spatiocyte \cite{ARJUNAN:2009SystSynthBiol} and the scheme developed by \person{Rigdway} et al. \cite{RIDGWAY:2008BiophysJ}, which use more accurate bimolecular interaction rates based on known analytical solutions; 
the latter further stands out by incorporating both fixed-step and event-driven time propagation.

\subsection{Detailed balance in BD simulations: Reaction Brownian Dynamics and the Reaction Volume Method}
\label{Sec-Sokolowski-RBD}
In this section we will explain two methods enabling BD simulations that obey detailed balance rigorously.  
When detailed balance holds, it is ensured that equilibrium properties of the system, 
such as state occupancies and radial distribution functions around reactants, are properly reproduced in simulations; 
the dynamics of the system is reproduced accurately on time scales of the propagation time step $\Delta t$ and larger.

Detailed balance demands that, for any distance vector $\myvec r_{\delta}$ between two interacting particles,
the probability $p_u(\myvec r_{\delta})d\myvec r_{\delta}$ of being in the unbound configuration (within an infinitesimal volume $d\myvec r_{\delta}$)
times the transition probability $\pi_{u\rightarrow b}(\myvec r_{\delta})$ to move into the bound state from $\myvec r_{\delta}$ 
equals the probability $p_b$ to be in the bound state times the probability of the inverse transition $\pi_{b\rightarrow u}(\myvec r_{\delta})$:
\begin{align}
  \label{Eq-Sokolowski-DetailedBalance}
  p_u(\myvec r_{\delta}) d\myvec r_{\delta} ~ \pi_{u\rightarrow b}(\myvec r_{\delta})  = p_b ~ \pi_{b\rightarrow u}(\myvec r_{\delta}) 
\end{align}
In situ, the state occupancy ratio $p_b/p_u(\myvec r_{\delta})=K_{eq}$ is fixed by the equilibrium constant of the reaction,
while the transition probability $\pi_{u\rightarrow b}(\myvec r_{\delta})$ depends on algorithmic details of particle propagation.
Thus, in order to impose detailed balance, one faces the task to prescribe a backward move in a way that $\pi_{b\rightarrow u}(\myvec r_{\delta})$
obeys Eq.~(\ref{Eq-Sokolowski-DetailedBalance}).

Reaction Brownian Dynamics (RBD) \cite{MORELLI:2008BiophysJ} makes use of analytical calculations to restore detailed balance in BD simulations that approximate particle contact by particle overlap.  
The crucial step herein is to rescale the reaction rate by a factor that correctly takes into account the statistics of spatial configurations from which reactions are attempted. 
For 3D diffusion, this factor can be computed exactly by integrating over all possible overlap situations generated by the diffusive motion of the particles. 
In a first step, the transition probabilities are rewritten as a product of a reaction proposal density, and a (reaction) acceptance probability: 
\newcommand{\Overlap}{\mathcal{O}}
\begin{align}
 \pi_{u\rightarrow b}(\myvec r_{\delta}) = P^{\rm
   gen}_{u\rightarrow\Overlap}(\myvec r_{\delta},\Delta t)P^{\rm
   acc}_{\Overlap\rightarrow b} (\Delta t)	\nn\\
 \pi_{b\rightarrow u}(\myvec r_{\delta}) = P^{\rm
   gen}_{\Overlap\rightarrow u}(\myvec r_{\delta},\Delta t)P^{\rm acc}_{b
   \rightarrow\Overlap} (\Delta t)\label{Eq-Sokolowski-ReactionProp}
\end{align}
Herein $\Overlap$ represents all overlap constallations from which reactions are allowed to occur, 
given by the spherical volume of radius $R$ equal to the sum of the particle radii, $R\equiv R_1 + R_2$. 
Combining Eqs.~(\ref{Eq-Sokolowski-DetailedBalance}) and (\ref{Eq-Sokolowski-ReactionProp}), one can further write
\begin{align}
 P^{\rm acc}_{\Overlap\rightarrow b} &= \frac{p_b}{p_u(\myvec r_{\delta})d\myvec{r_{\delta}}} 
					\frac{P^{\rm gen}_{\Overlap\rightarrow u}(\myvec r_{\delta},\Delta t)}{P^{\rm gen}_{u\rightarrow\Overlap}(\myvec r_{\delta},\Delta t)}
					P^{\rm acc}_{b \rightarrow\Overlap}
			= \frac{k_a}{k_d d\myvec{r_{\delta}}} \frac{P^{\rm gen}_{\Overlap\rightarrow u}(\myvec r_{\delta},\Delta t)}{P^{\rm gen}_{u\rightarrow\Overlap}(\myvec r_{\delta},\Delta t)} k_d \Delta t ~,
\end{align}
where in the last step we express the equilibrium constant $K_{eq}=p_b/p_u$ as the ratio of the intrinsic association rate $k_a$
and dissociation rate $k_d$, and exploit that the acceptance rate for the dissociation event is equal to $k_d\Delta t$. 
To obtain the acceptance rate for the association reaction, $P^{\rm acc}_{\Overlap\rightarrow b}$, 
we need the overlap generation probability for the association reaction, $P^{\rm gen}_{u\rightarrow\Overlap}(\myvec r_{\delta},\Delta t)$, 
and that for the dissociation reaction, $P^{\rm gen}_{\Overlap\rightarrow u}(\myvec r_{\delta},\Delta t)$. 
The former is given by 
\begin{align}
P^{\rm gen}_{u\rightarrow\Overlap}(\myvec r_{\delta},\Delta t) =\int_0^R dr_{\delta}^\prime (r_{\delta}^\prime)^2 \int_0^\pi d\theta \sin\theta \int_0^{2\pi}d\varphi ~ p(\textbf r_{\delta}^\prime,t+\Delta t|\textbf r_{\delta},t) ~,
\end{align}
where $\myvec{r}_{\delta}^\prime$ is a position inside the overlap sphere $\mathcal{O}$, i.e. $r_{\delta}^\prime = |\myvec{r}_{\delta}^\prime| \leq R$, and $p(\textbf r_{\delta} ^\prime,t+\Delta t|\textbf r_{\delta},t)$ is the Green's function for free diffusion\footnote{i.e., the probability (density) of finding the freely diffusing particle at position $\textbf r_{\delta}^\prime$ at time $t^\prime > t$, given that it started at position $\textbf r_{\delta}$ at time $t$}, given by Eq.~(\ref{Eq-Sokolowski-FreeGF}), with $\Delta\myvec{r}=\myvec{r}_{\delta}^\prime - \myvec{r}_{\delta}$; this integral can be performed analytically. 
The key idea of RBD is to take the generation distribution of the dissociation reaction, $P^{\rm gen}_{\Overlap\rightarrow u}(\myvec r_{\delta},\Delta t)$, to be equal to that of the association reaction, $P^{\rm gen}_{u\rightarrow\Overlap}(\myvec r_{\delta},\Delta t)$, but properly normalized: $P^{\rm gen}_{\Overlap\rightarrow u}(\myvec r_{\delta},\Delta t) \propto P^{\rm gen}_{u\rightarrow\Overlap}(\myvec r_{\delta},\Delta t)$.
This rule can be interpreted as a manifestation of the microscopic reversibility of diffusion trajectories.
More precisely, the distances of the dissociating particles are sampled from the distribution $P^{\rm gen}_{\Overlap\rightarrow u}(\myvec r_{\delta},\Delta t) d\myvec{\myvec r_{\delta}} \equiv \frac{1}{4\pi I_D(\Delta t)}P^{\rm gen}_{u\rightarrow\Overlap}(\myvec r_{\delta},\Delta t) d\myvec{\myvec r_{\delta}}$,
where $4\pi I_D(\Delta t)$ is a normalization factor which can be obtained by integrating $P^{\rm gen}_{u\rightarrow\Overlap}(\myvec r_{\delta},\Delta t)$ over all possible initial (target) positions for the forward (backward) move outside of $\mathcal{O}$:
\begin{align}
4\pi I_D(\Delta t) = \int_{|\myvec{r}|\geq R} P^{\rm gen}_{u\rightarrow\Overlap}(\myvec r_{\delta},\Delta t) d\myvec{r}_{\delta} = 4\pi \int_R^\infty P^{\rm gen}_{u\rightarrow\Overlap}(r_{\delta},\Delta t) r_{\delta}^2 dr_{\delta}
\end{align}
This ultimately yields the rescaled acceptance probability:
\begin{align}
 P^{\rm acc}_{\Overlap\rightarrow b} &= \frac{k_a \Delta t}{4\pi I_D(\Delta t)}
 \label{Eq-Sokolowski-RBDAccRate}
\end{align}
This result has a nice intuitive interpretation: the intrinsic association rate $k_a$ is the product of a collision frequency $4\pi I_D(\Delta t)/\Delta t$ times the probability $P^{\rm acc}_{\Overlap\rightarrow b}$ that a collision leads to association. 
The dominant contribution to the integral $I_D(\Delta t)$ comes from distances $r$ that are short compared to $\sqrt{D\Delta t}$. 
Hence, in the limit that $\Delta t \to 0$, the rate $k_a$ should approach the intrinsic association rate $k_a$ as used in the theories of diffusion-influenced reactions \cite{AGMON:1990JChemPhys};
here, $k_a$ is defined as the association rate given that the reactants are in perfect contact, i.e. touching but not overlapping.
We also note that this is the same definition of the intrinsic association rate as used in GFRD, discussed in Sec.~\ref{Sec-Sokolowski-GFRDAlgorithm}. 

In practice, detailed balance will be obeyed for any reasonable choice of $\Delta t$ if reaction attempts upon particle overlaps are accepted with probability $P^{\rm acc}_{\Overlap\rightarrow b}$, and if dissociating particles are placed with a randomly uniform angle at a radial distance $r\geq R$ sampled from the normalized distribution 
$P^{\rm gen}_{u\rightarrow \Overlap}(r,\Delta t)r^2/I_D(\Delta t)$; 
naturally, $\Delta t$ has to be chosen small enough such that $P^{\rm acc}_{\Overlap\rightarrow b}< 1$, meaning that $\Delta t$ again will be constrained by fast $k_a$ and $D$. 
Note that in RBD the reactants have a cross section that is determined by their physical size, not by their overall reaction rate; 
hence, reaction-limited reactions do not lead to small effective particle sizes and correspondingly small time steps, as in \Smoldyn. 
While RBD yields excellent results for diffusing spheres in 3D, it proved troublesome to extend the necessary analytical calculations to arbitrary dimensions and non-spherical objects.

A conceptually similar, but more versatile approach is the Reaction Volume Method \cite{PAIJMANS:2012MScThesis}.
Its key assumption is that reactive objects, be it particles or reactive surfaces, are surrounded by a small ``reaction volume'' $\mathcal{V}$
within which the precise shape of the density $p_u(\myvec r_{\delta})$ may be ignored.
This is equivalent to assuming that the probability density function of the interparticle separation is flat within the reaction volume, which is approximate but accurate if $\mathcal{V}$ is small with respect to the whole simulation space.
Reaction attempts only occur within $\mathcal{V}$, whereas actual overlap situations ($|\myvec{r_{\delta}}|<R$) are strictly rejected, and at the inverse reaction the particle is placed back into $\mathcal{V}$ uniformly.
The binding process is thus broken apart into a displacement and a reaction step;
as in the RBD algorithm, we can therefore split the transition probabilities in Eq.~(\ref{Eq-Sokolowski-DetailedBalance}) into two factors representing the proposal and acceptance probability, respectively:
\begin{align}
 \pi_{u\rightarrow b}(\myvec r_{\delta}) = P^{\rm gen}_{u\rightarrow \mathcal{V}}(\Delta t)P^{\rm acc}_{\mathcal{V} \rightarrow b}	\nn\\
 \pi_{b\rightarrow u}(\myvec r_{\delta}) = P^{\rm gen}_{\mathcal{V}\rightarrow u}(\Delta t)P^{\rm acc}_{b \rightarrow \mathcal{V}}
\end{align}
Again, similarly to RBD, it can be shown that these probabilities only differ by a factor, equal to $\mathcal{V}$ itself: $P^{\rm gen}_{u\rightarrow\mathcal{V}}(\Delta t) = \mathcal{V} P^{\rm gen}_{\mathcal{V}\rightarrow u}(\Delta t)$.
Once again assuming that dissociation events occur with Possonian statistics, i.e. $P^{\rm acc}_{b \rightarrow \mathcal{V}}=k_d\Delta t$,
one finds that detailed balance is fulfilled when forward reaction attempts are accepted with a rate
\begin{align}
 P^{\rm acc}_{\mathcal{V} \rightarrow b} = \frac{k_a\Delta t}{\mathcal{V}}	\quad .
 \label{Eq-Sokolowski-RVMAccRate}
\end{align}

The reaction volume $\mathcal{V}$ should be chosen as small as possible to minimize the error of the approximation made above.
On the other hand, in order to avoid high rejection rates,
$\mathcal{V}$ should be large enough to prevent particles from jumping over the reaction volume in just one time step; 
moreover, the requirement $P^{\rm acc}_{\mathcal{V}\rightarrow b}<1$ prohibits arbitrarily small $\mathcal{V}$ for fixed $k_a$ and $\Delta t$.
Optimal choices for $\mathcal{V}$ and $\Delta t$ thus are interdependent; in practice, it is advisable to set them simultaneously,
again taking into account the constraints imposed by the fastest reaction rates and diffusion coefficients.

\subsection{ReaDDy: Brownian dynamics with interaction forces}
\label{Sec-Sokolowski-ReaDDy}
Rooted in molecular dynamics approaches, a broad range of BD simulation environments taking into account hydrodynamic and electrostatic interactions emerged,
for example {\it UHBD} \cite{MADURA:1995ComputPhysCommun}, {\it Browndye} \cite{HUBER:2010ComputPhysCommun}, {\it BD\_BOX} \cite{DLUGOSZ:2011BMCBiophys} and {\it brownmove} \cite{GEYER:2011BMCBiophys}.
These schemes, however, can simulate only short time intervals ($\lesssim ms$), and---more im\-por\-tant\-ly---do not incorporate chemical reactions.
As an integrating approach, \ReaDDy \cite{SCHONEBERG:2013PLoSONE} explicitly aims at bridging the gap between particle-force interactions and chemical reaction kinetics.

\ReaDDy models the movement of particles with generalized \person{Langevin} dynamics in the overdamped limit\footnote{\person{Langevin} dynamics are conceptually introduced in Ch.~7; \person{Langevin} dynamics are called ``overdamped'' when the effects of forces acting on a particle are rapidly dissipated such that, on average, no acceleration takes place.}, and uses an \person{Euler} discretization algorithm to simulate the model. The underlying \person{Langevin} equation naturally incorporates an interaction potential (depending on the distance to other particles) via its drift term.
While in principle the potential energy may change continuously with changing particle positions, \ReaDDy makes the assumption that the potential remains
constant during the time between two updates.
To model bimolecular reactions, \ReaDDy labels particles that diffused to each other closely enough an ``encounter complex'' from which reactions can occur with an ``activation rate''. 
This, in essence, is akin to the approaches described in Sec.~\ref{Sec-Sokolowski-RBD}.
Since the rate with which the encounter complex is formed diffusively depends on the cut-off distance below which particles are considered ``encountered'',
the activation rate---to a certain extent---is artificial, and has to be tuned to match macroscopic equilibria;
\ReaDDy uses the rate derived by \person{Erban} and \person{Chapman} \cite{ERBAN:2009PhysBiol}, computed from the encounter radius, diffusion constants, and intrinsic reaction rate.
So far \ReaDDy does not yet strictly fulfill detailed balance, but ongoing work aims at ensuring this property in future versions.

\clearpage
\section{Event-driven schemes}
\label{Sec-Sokolowski-EventDriven}
As explained in the previous section, BD simulations require the propagation of each individual particle by small displacements at each update, while the accuracy of the simulation scales inversely with the average displacement length $\la |\Delta\myvec{r}| \ra$ and thus with the chosen fixed time step $\Delta t$.
This renders BD simulations particularly inefficient in ``sparse'' scenarios, i.e. at low particle density, when the average distance between two particles is much larger than their radii.
In such situations, a disproportionate amount of computational steps has to be spent on diffusing the particles before the truly interesting reaction events can even be attempted;
thus, an event-driven scheme---similar in spirit to the (non-spatial) \person{Gillespie} algorithm (SSA), but accounting for space---would be desireable.
In the following we describe two approaches to event-driven spatial-stochastic simulation that are conceptually different: First we briefly comment on RDME-based schemes which, in short, implement the \person{Gillespie} algorithm on spatial lattices; we then focus on truly particle-based schemes, which use analytical solutions to predict diffusive first-passage times and exit points in continuous space, taking into account chemical reactions between the particles.

 
\subsection{\person{Gillespie} algorithm on a lattice: RDME-based schemes}
\label{Sec-Sokolowski-RDME}
A broad range of schemes based on the Reaction-Diffusion Master Equation (RDME) extends the basic principle of the \person{Gillespie} algorithm (described in Ch.~7) to spatial lattices, modeling diffusive motion as hopping between neighboring voxels with concentration-dependent rates.
While RDME-based schemes can reach high computational efficiency, they require the existence of a length scale on which the system can be considered well-mixed, 
and thus do not reach the level of spatial detail achieved by particle-based schemes.
Overall, they are well-suited for somewhat higher concentrations at which stochasticity however cannot yet be neglected.

Prominent examples are: {\it MesoRD} \cite{HATTNE:2005Bioinform,FANGE:2006PLoSComputBiol,WANG:2013BullMathBiol}, implementing the Next-Subvolume-Method \cite{ELF:2004SystBiol}, {\it URDME} \cite{DRAWERT:2012BMCSystBiol} and associated techniques \cite{DRAWERT:2010JChemPhys,LAMPOUDI:2009JChemPhys,ISAACSON:2006SIAMJSciComput} (which recently lead to the development of {\it StochSS} \cite{DRAWERT:2016PLoSComputBiol}), {\it VCell} \cite{MORARU:2008IETSystBiol} and {\it GMP} \cite{RODRIGUEZ:2006Bioinform,VIGELIUS:2010Bioinform}; also spatial tau-leaping techniques \cite{IYENGAR:2010JChemPhys} belong to this group.
For methodic details we refer the reader to Ch.~22, which is particularly devoted to RDME-based approaches.
However, in Sec.~\ref{Sec-Sokolowski-Hybrid} we describe the recently devised Small-Voxel Tracking Algorithm (\SVTA) \cite{GILLESPIE:2014JChemPhys}.
While using a lattice and thus superficially similar to RDME-based schemes, the \SVTA fundamentally differs from these in that it explicitly simulates the diffusion and reactions of individual particles, borrowing ideas from the \eGFRD scheme described in the following section.

\subsection{Particle-based event-driven schemes: GFRD and FPKMC}
\label{Sec-Sokolowski-GFRD-FPKMC}
Green's Function Reaction Dynamics (\GFRD) is an exact event-driven algorithm for simulating reaction-diffusion systems at the particle level \cite{VAN_ZON:2005PRL,VAN_ZON:2005JChemPhys,TAKAHASHI:2010PNAS}. 
A reaction-diffusion system is a many-body problem that cannot be solved analytically. 
The original idea of \GFRD is to decompose the many-body problem into one- and two-body problems that can be solved analytically via Green's functions, 
and to use these Green's functions to set up an event-driven algorithm \cite{VAN_ZON:2005PRL,VAN_ZON:2005JChemPhys}. 
The Green's functions allow \GFRD to make large jumps in time and space when the particles are far apart from each other. 
Indeed, under biologically relevant conditions, corresponding to ${\rm nM}-\mu{\rm M}$ reactant concentrations, \GFRD is $4-6$ orders of magnitude more efficient than brute-force Brownian Dynamics\footnote{For more information see: \texttt{http://gfrd.org}}.

In the original version of the algorithm, the many-body problem was solved by determining at each update of the simulation a maximum time step such that each particle could interact with at most one other particle during that time step \cite{VAN_ZON:2005PRL,VAN_ZON:2005JChemPhys}. 
This scheme was a synchronous event-driven algorithm, because at each update all the particles were propagated simultaneously. 
Moreover, the scheme was not exact, because the decomposition into single particles and particle pairs involved cut-off distances, introducing a trade-off between speed and error.

A newer version of the algorithm, called \eGFRD, \cite{TAKAHASHI:2010PNAS}, implements the idea of \textit{protective domains}, originally introduced by \person{Oppelstrup} and co-workers \cite{OPPELSTRUP:2006PRL}.
In \eGFRD, protective domains of simple geometric shape are put around single particles and pairs of particles. 
For each of the domains, an exact analytical solution to the reaction-diffusion problem is computed using Green's functions. 
This yields for each domain an {\it event type} and an {\it event time;} 
as described below, the set of possible event types depends on whether the domain is a {\it Single}, meaning that it contains a single particle, or a {\it Pair}, containing a pair of particles. The event times are collected in a chronologically ordered event list, and the events are then executed in chronological order. 
When an event is executed, first the particles of the corresponding domain are propagated;
then new domains, with new event types and new event times, are determined for the propagated particles, and the new events are put back into the event list. 
\eGFRD, is thus an exact, event-driven, asynchronous algorithm. 
The use of protective domains, and their asynchronous updating, makes \eGFRD not only exact, but also faster than the original \GFRD scheme. 
More recently, \eGFRD has been extended to 1D and 2D \cite{SOKOLOWSKI:2013PhDThesis,PAIJMANS:2014PhysRevE,WEHRENS:2014JChemPhys}, which makes it possible to simulate reactions and diffusion at membranes \cite{WEHRENS:2014JChemPhys}, reactions and diffusion along the DNA \cite{PAIJMANS:2014PhysRevE}, as well as active transport along cytoskeletal filaments \cite{SOKOLOWSKI:2013PhDThesis}.

In parallel, \person{Oppelstrup} {\it et al.} also developed an asynchronous, event-driven reaction-diffusion algorithm, called First-Passage Kinetic Monte Carlo (\FPKMC) \cite{OPPELSTRUP:2009PhysRevE}. 
In contrast to \eGFRD, it assumes the reactions to be diffusion limited. 
Moreover, the reactants are modeled via little cubes, instead of spheres as in \eGFRD. 
Their asynchronous nature make \eGFRD and \FPKMC similar in spirit to event-driven MD simulations of hard spheres and the \person{Gibson}-\person{Bruck} scheme, which is an exact, event-driven, asynchronous algorithm for simulating the zero-dimensional chemical master equation \cite{GIBSON:2000JPhysChemA} (see Ch.~7).

Due to the close conceptual similarity of \eGFRD and \FPKMC, here we only discuss \eGFRD in detail; 
for recent developments of the \FPKMC method, we refer the interested reader to the respective literature \cite{SCHWARZ:2013JComputPhys,MAURO:2014JComputPhys}.
Below, we first explain the analytical basis of \eGFRD, i.e. how Green's functions can be obtained mathematically.
We then describe how the Green's functions can be used to implement an event-driven spatial-stochastic simulation algorithm. 
For more details we refer to \cite{SOKOLOWSKI:2013PhDThesis}, which gives not only a detailed description of the algorithm, but also explains the derivation of the necessary Green's functions,
in particular for reactions and diffusion in 1D and 2D, not discussed here.

\subsubsection{Mathematical basis: \person{Smoluchowski} equation and PDE solutions}
\label{Sec-Sokolowski-GFRDMaths}
\begin{figure}[t]
  \centering
  \includegraphics[width=\textwidth]{\FigDir/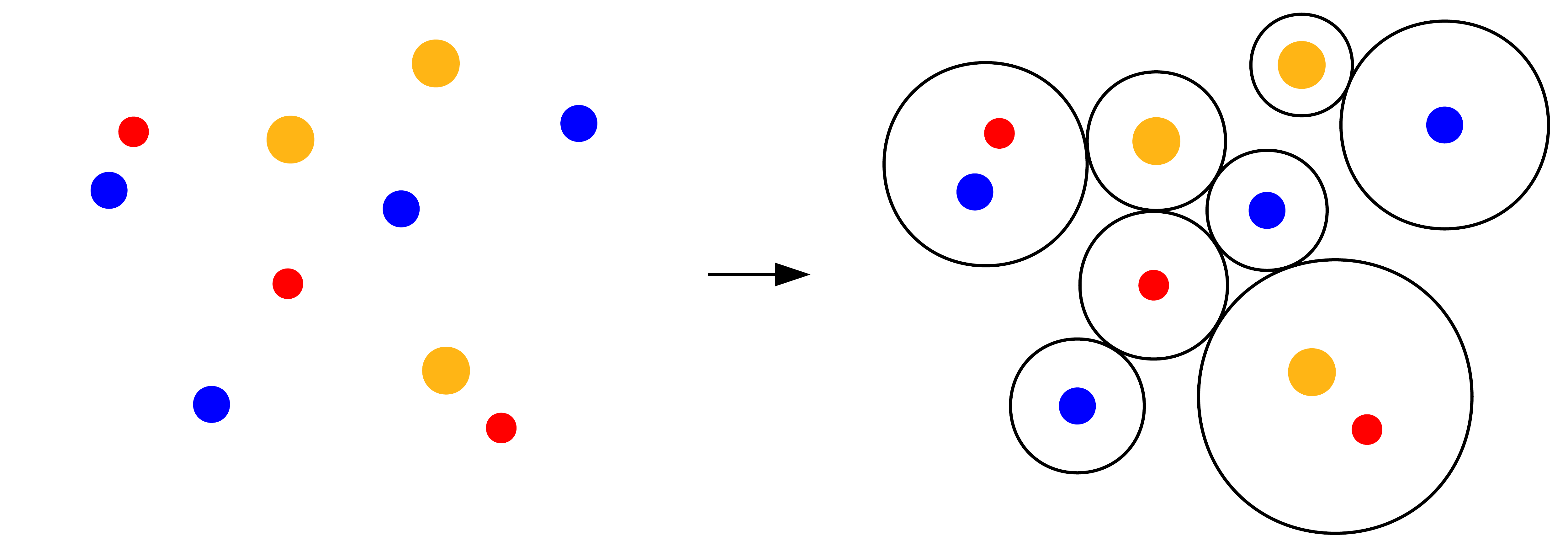}  
\caption{ \label{Fig-Sokolowski-GFRD-Principle}
  \textbf{In \eGFRD protective domains separate the $N$-particle problem into one- and two-particle problems.}
	  The drawing illustrates how \eGFRD constructs protective domains that contain at most two particles
	  in order to isolate these from the influence of other particles,
	  starting from a random spatial configuration of the particles.
	  Subsequently, analytical solutions are calculated for each domain individually and used to
	  propagate the domains in an event-driven, asynchronous fashion.
	  We show here a 2D projection for the standard scenario in which particles diffuse and react in 3D space.
	  In this case protective domains are spherical.
	  Different colors mark different chemical species.
}
\end{figure}

Consider a random spatial configuration of $N$ particles representing a typical situation in the model
defined in Sec.~\ref{Sec-Sokolowski-Intro}, which assumes that the solid spherical particles 
are completely characterized by their radii, diffusion constants, and rates of mutual interaction and decay (cf. Fig.~\ref{figStdModel}).
Even with these simplifications, in general it is hard---if not impossible---to find an analytical prediction 
for future particle positions and species given that the system started from a certain initial condition.
Nonetheless, as often in physics, exact analytical solutions can be obtained for the case $N\leq2$.
\eGFRD capitalizes on this fact by dividing the 3D volume into subvolumes, called \textit{protective domains},
that contain at most two particles, in order to isolate the content of each domain from
the influence of surrounding particles up to a certain (domain-specific) time $\tau_{\mathcal D}$.
This way the $N$-particle problem is reduced to $M<N$ independent one- or two-particle problems.
$\tau_{\mathcal D}$ is the time at which a reconstruction of the domain becomes necessary,
e.g. when one of the particles hits a domain boundary or experiences a reaction that changes its properties.
Fig.~\ref{Fig-Sokolowski-GFRD-Principle} illustrates this principle.
Here the domains that enclose the single particles and particle pairs, which in the following we will call \textit{Single} and \textit{Pair} domains, respectively, are spherical, but they could be of any geometric type.

For sufficiently simple domain geometries, in particular spheres or cylinders, the Green's functions for the isolated reaction-diffusion problems, i.e. the density function $p(\myvec r,t|\myvec r_0)$ for the probability that a particle is at position $\myvec r$ at time $t$ given that it started at position $\myvec r_0$, can be calculated analytically with exact results.
Here the confining character of the domain is taken into account by imposing specific boundary conditions to $p(\myvec r,t|\myvec r_0)$.

Let us first consider the \textit{Single} domain case.
Here, the dynamics of the diffusing particle inside the domain is captured by the diffusion equation
\begin{align}
 \ddt p(\myvec r,t|\myvec r_0) = D \nabla_{\myvec r}^2 p(\myvec r,t|\myvec r_0) + \delta(\myvec r - \myvec r_0)\delta(t - t_0)~,
 \label{Eq-Sokolowski-DiffSingle}
\end{align}
where $D$ the diffusion constant, and $\nabla_{\myvec r}^2 \equiv \frac{\partial^2}{\partial x^2} + \frac{\partial^2}{\partial y^2} + \frac{\partial^2}{\partial z^2}$ for $\myvec{r} = (x,y,z)^T$.
Note that due to the delta-peak inhomogeneity that represents the initial condition,
the solution $p(\myvec r,t|\myvec r_0)$ technically indeed is a Green's function.
To sample a first-passage time for the particle to reach the outer shell $\partial\mathcal{D}_1$
of a domain $\mathcal{D}_1$ constructed around $\myvec r_0$,
additionally one must impose the following absorbing boundary condition:
\begin{align}
 p(\myvec r,t|\myvec r_0) = 0 \quad\text{for}\quad \myvec r \in \partial\mathcal{D}_1
 \label{Eq-Sokolowski-BCSingle}
\end{align}
In the simplest case, for a spherical domain with radius $R$, this is equivalent to:
\begin{align}
 p(|\myvec r - \myvec r_0|=R,t|\myvec r_0) = 0
 \label{Eq-Sokolowski-BCSingleSphere}
\end{align}
In spherical coordinates, the boundary value problem defined by Eqs.~(\ref{Eq-Sokolowski-DiffSingle}) and (\ref{Eq-Sokolowski-BCSingleSphere}) can be solved exactly via standard methods, 
such as eigenfunction expansion or Laplace transforms. 
For more complicated domain geometries, e.g. cylinders, the mathematical problem has to be transformed into a coordinate system that captures specific symmetries, 
and boundary conditions have to be imposed for each coordinate separately.

\begin{figure}[t]
  \centering
  \includegraphics[width=0.4\textwidth]{\FigDir/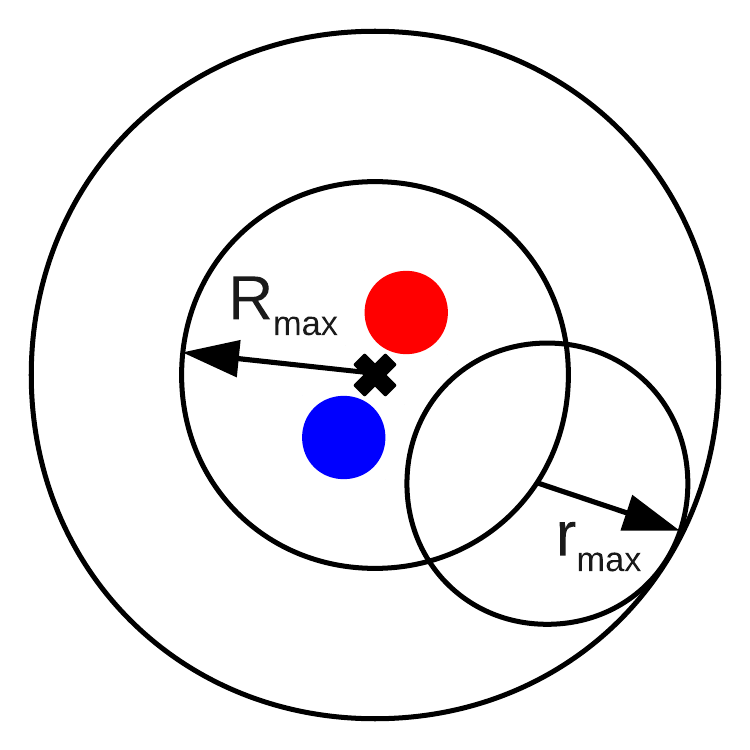}  
\caption{ \label{Fig-Sokolowski-GFRD-Pair}
  \textbf{Decomposition of a \textit{Pair} domain into subdomains.}
  The sketch shows a valid decomposition of a \textit{Pair} domain in \eGFRD into subdomains for the weighted center-of-mass vector $\myvec{R}$
  and the interparticle vector $\myvec{r}$.
  The case shown applies to spherical \textit{Pair} domains in 3D or circular \textit{Pair} domains in 2D.
}
\end{figure}

The analogous calculations for a \textit{Pair} domain $\mathcal{D}_2$ follow the same principles as for \textit{Singles}.
However, here the two particles can react at contact, which creates an additional exit channel and corresponding next-event type. 
Importantly, the problem of two particles that diffuse in a bounded domain and that can also react with each other upon contact, cannot be solved directly using Green's functions. 
Here \eGFRD employs a trick, in which the original problem is decomposed into one diffusion problem for the center-of-mass and one reaction-diffusion problem for the interparticle vector of the two particles. 
Let us denote by $p_2(\myvec r_A, \myvec r_B, t|\myvec r_{A0}, \myvec r_{B0})$ the
probability density function for the likelihood of finding two
diffusing particles $A$ and $B$, initially located at positions
$\myvec r_{A0}$ and $\myvec r_{B0}$ at $t=t_0$, at positions $\myvec
r_{A}$ and $\myvec r_{B}$ at a later time $t$.  The time evolution of
$p_2$ is governed by the \person{Smoluchowski} equation:
\begin{align}
 \label{Eq-Sokolowski-Pair}
 \ddt p_2(\myvec r_A, \myvec r_B, t|\myvec r_{A0}, \myvec r_{B0})
    &=	\left[ D_A\nabla_{\myvec{r}_A}^2 + D_B\nabla_{\myvec{r}_B}^2 \right] ~ p_2(\myvec r_A, \myvec r_B, t|\myvec r_{A0}, \myvec r_{B0})
\end{align}
Here $D_A$ and $D_B$ are the diffusion constants of particles $A$ and $B$.
This problem can be simplified 
by transforming coordinates $\myvec r_A$ and $\myvec r_B$ to $\myvec r$ and $\myvec R$,
where $\myvec r\equiv\myvec r_B - \myvec r_A$ is the interparticle vector 
and $\myvec R\equiv (D_B \myvec r_A + D_A \myvec r_B) / (D_A+D_B)$ is a (weighted) center-of-mass of the particles.
A separation ansatz $p_2=p_r(\myvec r)p_R(\myvec R)$
then yields two separate, uncoupled diffusion equations for $\myvec r$ and $\myvec R$,
which are equivalent to (\ref{Eq-Sokolowski-Pair}):
\begin{align}
 \ddt p_r(\myvec r, t|\myvec r_0) = D_r \nabla_{\myvec r}^2\, p_r(\myvec r, t|\myvec r_0)\;, & & \ddt p_R(\myvec R, t|\myvec R_0) = D_R \nabla_{\myvec R}^2\, p_R(\myvec R, t|\myvec R_0) \quad .
 \label{Eq-Sokolowski-PairSeparated}
\end{align}
Herein, $D_r\equiv D_A + D_B$ and $D_R\equiv D_A D_B / (D_A+D_B)$ \footnote{
More generally, it can be shown that there is some freedom in defining the coordinate transform
and the resulting diffusion constants for the transformed coordinates, depending on the precise definition of $\myvec R$ \cite[p.42ff]{SOKOLOWSKI:2013PhDThesis}.
}.
The uncoupling allows for the calculation of two Green's function solutions
$p_r(\myvec r,t|\myvec r_0)$ and $p_R(\myvec R,t|\myvec R_0)$ on two subdomains
$\mathcal{D}_r$ and $\mathcal{D}_R$ of $\mathcal{D}_2$, respectively, with
boundary conditions adapted to the problem as described further below.
$\mathcal{D}_r$ and $\mathcal{D}_R$ must be defined in a way that 
all possible positions constructed from sampled values of $\myvec r$ and $\myvec R$
remain within the protective domain $\mathcal{D}_2$.
Fig.~\ref{Fig-Sokolowski-GFRD-Pair} shows a valid 
definition of the subdomains for a (projected) spherical pair domain.

The separation into center-of-mass $\myvec R$ and interparticle vector
$\myvec r$ allows us to model reactions between $A$ and $B$ 
by imposing a radiating boundary condition to $p_r$ at the particle contact radius $\sigma = R_A + R_B$ as follows:
\begin{align}
 \int_{\partial\mathcal{D}_r^\sigma} -D_r\nabla_{\myvec r}p_r(\myvec r,t|\myvec r_0) d\myvec r = k_a p_r(|\myvec r| = \sigma, t)
 \label{Eq-Sokolowski-BCPair}
\end{align}
Here, $k_a$ is the intrinsic particle reaction rate, which is the rate
at which the particles react given that they are in contact, and
$p_r(|\myvec r| = \sigma, t)$ is the probability that the particles
are indeed at contact at time $t$.  The integral on the left is the
total probability (out)flux through the ``contact surface'' or inner
boundary of the $\myvec r$-subdomain, which is the set of all points
at which $A$ and $B$ are in contact: $\partial\mathcal{D}_r^\sigma =
\left\lbrace \myvec r \big\vert |\myvec r| = \sigma \right\rbrace$.
At the outer boundary of the $\myvec r$-subdomain,
$\partial\mathcal{D}_r^a$, absorbing boundary conditions are imposed.
The initial condition for this boundary value problem is set by the
inital separation of the two particles, $p_r(\myvec r, t=0|\myvec
r_0)=\delta(\myvec r - (\myvec r_{B0} - \myvec r_{A0}))$.  While the
boundary problem defined by
Eqs.~(\ref{Eq-Sokolowski-PairSeparated}),(\ref{Eq-Sokolowski-BCPair}) and (\ref{Eq-Sokolowski-BCSingle})
(with $\partial\mathcal{D}_1=\partial\mathcal{D}_r^a$) is more
complicated than in the \textit{Single} domain case, also here an
exact solution for $p_{\myvec r}$ can be obtained for sufficiently
simple geometries.  The problem for $p_{\myvec R}$ is solved precisely
in the same way as for the \textit{Single} domain.

Since the form of the Laplacian varies accross different
dimensions, the Green's functions for reaction-diffusion in lower
dimensions have to be rederived for these geometries, together
with the quantities that follow from them, needed for sampling the next event times and types 
(described in Sec.~\ref{Sec-Sokolowski-GFRDAlgorithm}).  Largely based on
the theory of heat conduction, this has been accomplished for the
2D-reaction-diffusion and 1D-reaction-diffusion-convection processes,
and even for the case in which two interacting particles diffuse in 2D
and 3D, respectively \cite{SOKOLOWSKI:2013PhDThesis}.  The derived Green's
functions then can also be used to sample event times for interactions
of particles with flat and cylindrical reactive surfaces,
representative of the cell membrane or intracellular filaments such as
cytoskeletal tracks or DNA.  Based on the geometry of the specific
situation considered, in the extended, all-dimensional version of
\eGFRD the protective domains are either spherical or cylindrical
\cite{SOKOLOWSKI:2013PhDThesis}.

The Green's functions thus differ depending on which particular situation they describe, such that it is beyond the scope of this chapter to discuss them in detail; 
we therefore limit ourselves to presenting their general structure.
An overview of the Green's functions derived and implemented in \eGFRD can be found online\footnote{See: \texttt{https://github.com/gfrd/egfrd/tree/develop/doc/greens\_functions}}.
Mathematically, the Green's functions for the cases in which a particle interacts with another reactive object (such as a second particle)
represent initial value problems that are double-bounded in space;
we thus expect them to take the form of infinite sums over the corresponding spatial eigenfunctions, weighted by a time-dependent factor.
In accordance, in the considered cases the Green's functions follow the form
\begin{align}
 p_r(r,t|r_0) \sim \sum_n e^{- \rho_n^2(k_a,D) t} \mathcal{F}_n(r) \mathcal{F}_n(r_0)
\end{align}
where $\rho_n(k_a,D)$ are the roots of an (oftentimes implicit) equation
that depends on the intrinsic reaction rate $k_a$ and diffusion constant $D$.
In the higher dimensions, especially in 2D, the eigenspectrum can be more complicated than in the case above and may require more than one summation.
The eigenfunctions $\mathcal{F}_n$ oftentimes take nontrivial forms, involving special functions.
In practice, when the values of the Green's function and derived quantities are computed numerically,
the infinite summation has to be truncated according to reasonable convergence criteria.
Since the exponential prefactor usually scales $\sim n^2$, the functions converge rather quickly,
but this behavior increasingly breaks down for very small times $t$.
Importantly, even when numerical approximations have to be made to compute the Green's functions,
their accuracy can always be increased by higher investment of computational resources,
while the \eGFRD scheme as such remains exact throughout.

\subsubsection{Algorithmic details of \eGFRD}
\label{Sec-Sokolowski-GFRDAlgorithm}
Quantities that derive from the Green's function $p(\myvec r,t|\myvec r_0)$ can be used to generate tentative
next-event times for each domain individually.
If collected in a global scheduler list, the sampled times can be used to update the domains
sequentially (i.e. asynchronously) and to set up an event-driven scheme.
While updates result in particle displacements and possibly species changes,
by construction these remain confined to the respective domain and thus do not interfere
with the situation in neighboring domains.

How can we sample next-event times and update particle positions using the Green's function $p(\myvec r,t|\myvec r_0)$?
Integration of $p(\myvec r,t|\myvec r_0)$ over its mathematical domain of support $\mathcal{D}$\footnote{For a spherical \textit{Single} domain, $\mathcal{D}$ is equal to the volume of the actual protective domain in which the radius is reduced by the particle radius.} 
yields the survival probability $S(t)$, i.e. the probability for the particle(s) to still remain within $\mathcal{D}$ at time~$t$.
Note that $S(t_0)=1$ given that $t_0$ is the domain construction time.
The survival probability is linked to the propensity function $q(t)$, which is the
probability for exiting through (any part of) the domain boundary $\partial\mathcal{D}$ within the time interval $[t,t+dt]$, via:
\begin{align}
  q(t) = -\ddt S(t) = -\ddt \int_{\mathcal{D}} p(\myvec r, t|\myvec r_0) d\myvec r
\end{align}
In other words, $1-S(t) = \int_{t_0}^t q(t')dt' = Q(t)$ is equal to the 
cumulative distribution function of $q(t)$ and may be used to sample 
a next-event time $\tau_e$ for leaving the domain 
via the inversion method as follows:
\begin{align}
 \tau_e = Q^{-1}(\mathcal{R}_e) = S^{-1}(1-\mathcal{R}_e)
\end{align}
Here $\mathcal{R}_e\in[0,1]$ is a uniformely distributed random number.
In general, it is difficult to calculate $S^{-1}$ analytically.
Then $\tau_e$ can be obtained by solving the equation $S(\tau_e) - \mathcal{R}_e =0$
with a numerical rootfinder\footnote{As a matter of course, using $1-\mathcal{R}_e$ and $\mathcal{R}_e$ is
equivalent if both are uniform random numbers from $[0,1]$.}.

In \textit{Pair} domains, the weighted center of mass $\myvec{R}$ and the interparticle vector $\myvec{r}$ are both diffusing 
independently on their respective domains of support; here we therefore obtain two tentative exit times $\tau_R$ and $\tau_r$,
and whichever of them is smaller decides on whether the sampled event is an escape of the $\myvec{R}$ or $\myvec{r}$ coordinate.
For the \textit{Single} domain, there is only one way of exiting, namely by hitting the outer, absorbing domain boundary;
this also holds for the weighted center-of-mass $\myvec{R}$ in \textit{Pair} domains.
In contrast, the interparticle vector $\myvec{r}$ in \textit{Pair} domains can escape through two boundaries: the reactive inner boundary at $|\myvec{r}|=\sigma$, where $\sigma$ is the particle contact radius, or through the outer absorbing boundary. The exit through the inner boundary corresponds to a reaction event.
When $\tau_r<\tau_R$, meaning that the event scheduled to happen in a \textit{Pair} domain indeed is an escape in $\myvec{r}$, it is determined from the relative magnitude of the probability fluxes $q_\sigma(\tau_r)$ and $q_a(\tau_r)$ through the respective boundaries at the event time $\tau_r$ whether the event is a reaction, or an escape through the outer boundary.

In addition to the exit events described above, in principle any particle can undergo a unimolecular reaction that results in a species change, annihilation, or dissociation into two educts; 
assuming Poissonian statistics, the respective event times $\tau_u$ can be sampled by equating the cumulative function $P(t) = 1 - e^{-k_u t}$ with a uniform random number $\mathcal{R}_u \in[0,1]$ and inverting for $t$, where $k_u$ is the corresponding unimolecular reaction rate\footnote{In case of several independent unimolecular reaction channels with rates $k_j$, it is possible to sample $\tau_u$ in the same way with $k_u=\sum_j k_j$, determining the actual reaction to occur from the fractional propensities in a second step, as in the \person{Gillespie} algorithm.}.
Overall, we thus have to sample (up to) two event times ($\tau_e$, $\tau_u$) in the \textit{Single} domain and (up to) four event times ($\tau_r$, $\tau_R$, $\tau_{u1}$, $\tau_{u2}$\footnote{Within \textit{Pair} domains, unimolecular reactions can happen for both contained particles separately.}) in the \textit{Pair} domain, respectively; 
in both cases, the minimal time decides for the type of the event to happen.
Note that the event time and type do not yet determine the new particle position(s);
however, the complete knowledge of the time evolution of the spatial probability density inside the domains allows us to sample particle positions at arbitrary event times $\tau$,
via the cumulative function ($\myvec r$-integral) of $p(\myvec r,\tau|\myvec r_0)$, again using the inversion method described above.

Since in \eGFRD the advance in simulated time and distance directly
correlates with the size of a protective domain, ideally one should
make them as large as possible (unless they would take away too much
space for neighboring domains).  However, due to the requirement of
strictly avoiding domain overlaps, in a concrete simulation nearby
obstacles such as other domains or reactive surfaces will set hard
limits to the maximal domain size. Constructing domains of
ever decreasing size---while technically possible down to the limit
set by the particle radius---would constitute a waste of computational
resources, because the sampling of next-event times from the Green's
function is computationally expensive; obviously, there is a minimal
domain size below which even naive Brownian dynamics (BD) schemes
(cf. Sec.~\ref{Sec-Sokolowski-BD}) become more efficient in propagating the
particle.  In crowded situations, \eGFRD thus uses BD as a
``fallback'' system. In order to allow particles to seamlessly
transfer between the \eGFRD and BD algorithm, \eGFRD makes use of
special {\it Multi} domains within which particles are propagated via BD. 
While overlaps are strictly forbidden for the regular protective domains of \eGFRD,
{\it Multi} domains can overlap with each other and with reactive surfaces,
allowing for particle encounters and reaction attempts upon BD displacements.
When particles enclosed in Multi domains have diffused sufficiently far away from the obstacles that prompted them to transfer to BD mode, 
the construction of regular protective domains around them can resume. The critical distances at which particles change their propagation mode---BD or GFRD---are tuned via length scales that are optimization parameters in the simulations. 
Multi domains have a trivial next-event time equal to the propagation step $\Delta t$ of the BD simulation running inside them, which can be conveniently included into the global scheduler of \eGFRD.

Another feature of \eGFRD that is critically important for the
creation of \textit{Pair} domains, but also useful in preventing
overly small domain sizes, is ``domain bursting'': Whenever a particle
updated at time $t$ ends up close to an already existing domain with
scheduled update time $t'$, it can ``burst'' the exising domain,
i.e. cause its premature update at $t<t'$. This prompts the removal
of the existing domain and update of its particle position(s), resulting
in a potentially more favorable situation for making domains at time
$t$ that possibly allows for the creation of a \textit{Pair} domain
around two particles previously contained in different domains. Here,
again, the length scale setting the criterion for being ``close'' is
tuned by a simulation parameter.

Algorithm~\ref{Alg-Sokolowski-eGFRD} summarizes the basic algorithm of \eGFRD in a logical listing;
note that here the rules for resorting to Brownian dynamics described above are implicitly contained in the rules for domain making, valid for both regular and Multi domains.

\begin{algorithm}
\begin{algorithmic}[p!]
 \vspace{1EM}
 \State Initialize:
 \State $t_{\rm sim} \gets 0$, scheduler $S \gets \lbrace\rbrace$
 \ForAll{ particles $p_i$ }
      \If{ \NOT $p_i$ already in domain }
	\State $\mathcal{D}_j \gets$ create domain for $p_i$
	\State $\tau_j \gets$ draw next-event time for $\mathcal{D}_j$
	\State insert $\tau_j$ into $S$ ordered by increasing time
      \EndIf
 \EndFor
 \State
 \State Main loop:
 \While{ $S \neq \lbrace\rbrace$ \AND $t_{\rm sim} < t_{\rm end}$}
    \State $t_{\rm sim} \gets \tau_n =$ topmost element in $S$
    \State remove $\tau_n$ from $S$
    \State propagate $\mathcal{D}_n$ to $\tau_n$ and remove $\mathcal{D}_n$
    \State reset particle update list: $U \gets \lbrace\rbrace$
    \State $U \gets U \cup \lbrace p_{n_i} \rbrace$ \textbf{for all} particles $p_{n_i} \in \mathcal{D}_n$
    \While{ $U \neq \lbrace\rbrace$ }
	\State $p_u \gets$ next particle in $U$
	\ForAll{domains $\mathcal{D}_{u_j}$ close to $p_u$ }
	    \State burst: propagate $\mathcal{D}_{u_j}$ to $\tau_n$ and remove $\mathcal{D}_{u_j}$
	    \State remove $\tau_{u_j}$ from $S$	    
	    \State $U \gets U \cup \lbrace p_{u_{jk}} \rbrace$ \textbf{for all} particles $p_{u_{jk}} \in \mathcal{D}_{u_j}$
	\EndFor
    \EndWhile
    \ForAll{ $p_u \in U$ }
	\If{ \NOT $p_u$ already in domain }
	    \State $\mathcal{D}_u \gets$ create domain for $p_u$
	    \State $\tau_u \gets$ draw next-event time for $\mathcal{D}_u$
	    \State insert $\tau_u$ into $S$
	\EndIf
    \EndFor    
 \EndWhile
 \vspace{1EM}
\end{algorithmic}
\caption{
Basic outline of the \eGFRD algorithm.
Symbols $\mathcal{D}_x$ denote domains, $\tau_x$ next-event times.
The scheduler $S$ is the list of all next-event times in the system, ordered by increasing time.
List $U$ collects all particles that have been updated at a given time $\tau_x$ and require construction of a new domain.
$t_{\rm sim}$ is the time that passed since simulation start.
\label{Alg-Sokolowski-eGFRD}
}
\end{algorithm}

\pagebreak
\section{Recent developments: Hybrid schemes and parallelization}
\label{Sec-Sokolowski-Hybrid}
As outlined in Sec.~\ref{Sec-Sokolowski-Intro}, the spatial detail necessary to capture relevant effects in spatial-stochastic simulations and the computational performance of the respective schemes strongly depend on the (local) density of the simulated particle crowd. Moreover, the abstract model introduced in Sec.~\ref{Sec-Sokolowski-Intro}, which serves as a basis for most common algorithms, ignores the internal structure of interacting particles that often is relevant to their chemical behavior. The awareness of these limitations recently has driven the development of various hybrid and multiscale schemes that aim at bridging simulation algorithms with different degrees of detail, such that propagation modes can be seamlessly switched in a situation-dependent manner.

Ongoing efforts include the elaboration of methods that---within one simulation---correctly transfer particles between regions of high particle density favoring PDE-based modeling, regions of intermediate density eligible for lattice-based simulation, and regions of such low density that genuinely particle-based schemes become indispensable \cite{ERBAN:2013BullMathBiol,FRANZ:2013SIAMJApplMath,FLEGG:2014SIAMJSciComput,FLEGG:2015JComputPhys,ROBINSON:2014JChemPhys}. 
Here, one practical difficulty is the ``conversion'' of continuous probability density leaking out of PDE-modeled regions into discrete particle numbers, which are always well-defined in the two other regimes---a naive treatment of the movement into and out of PDE-modeled regions can lead to artificial creation and annihilation of particles at the region interface.
As a concrete application of the theories coupling particle- and lattice-based algorithms, recently a multiscale version of \Smoldyn has been developed \cite{ROBINSON:2015Bioinform}.

For \eGFRD, recent work aims at an even more accurate modeling of the
reaction process. In MD-GFRD (Molecular Dynamics GFRD)
\cite{VIJAYKUMAR:2015JChemPhys}, particles are no longer treated as ideal
spheres when coming close to each other, but rather are propagated via
schemes that make it possible to resolve their detailed structure; depending on
the required level of detail, either \person{Langevin} dynamics or
\person{Markov} state models obtained from molecular dynamics
simulations are used to simulate particles in proximity, or
``contact''.  Conceptually similar ongoing work pursues the
integration of \eGFRD into \ReaDDy (cf. Sec.~\ref{Sec-Sokolowski-ReaDDy}).
Moreover, as part of the \eCell project
\cite{TOMITA:1999Bioinform,TAKAHASHI:2003Bioinform,TAKAHASHI:2004Bioinform,YACHIE-KINOSHITA:2010BioMedResInt},
which unifies a broad range of techniques under the ambitious
long-term goal to enable stochastic simulation of a whole cell,
recently a parallel version of \eGFRD, labeled \pGFRD, and a lattice-based
event-driven reaction-diffusion simulation scheme named \pSpatiocyte
have been devised \cite{MIYAUCHI:2016arXiv}; note that given their intrinsically asynchronous
nature, parallelization of event-driven schemes requires smart
treatment of particle transfers between independent, possibly strongly
desynchronized subregions of the simulated space.

We end by briefly describing the Small Voxel Tracking Algorithm
(\SVTA), recently developed by \person{Gillespie} and co-workers
\cite{GILLESPIE:2014JChemPhys}. The \SVTA is particle-based and driven by the
same spirit as \eGFRD in that it partitions the simulation space into
domains that contain at most two-particles. However, unlike in \eGFRD,
the event times for exiting the domains are not sampled from exact
analytical functions, but generated via a voxel-hopping algorithm that
simulates particle diffusion inside the domains on a lattice in an
asymptotically exact way. This obviously demands the voxels to be
smaller than the domains. In fact, in order to simulate bimolecular
reactions accurately, in the \SVTA the voxels are required to be even
smaller than the particles; the reaction times then are determined by
detailed bookkeeping of random particle collisions on the voxel grid,
grounded in a rigorous microscopic theory.

\section{Further reading}
For a more detailed comparison of different particle-based stochastic simulation schemes we refer the interested reader to several reviews in the literature \cite{DOBRZYNSKI:2007Bioinform,BURRAGE:2011Book,KLANN:2012IntJMolSci}.
We also recommend the practical guide by \person{Erban} and co-workers \cite{ERBAN:2007arXiv}.
A more detailed introduction into \eGFRD can be found in \cite{SOKOLOWSKI:2013PhDThesis};
detailed descriptions of mathematical techniques for deriving Green's functions (of the diffusion/heat equation) 
can be found in the literature on the theory of heat conduction \cite{CARSLAW_JAEGER:1959Book,OZISIK:2002Book,BECK:2010Book}.

\section{Online resources}
Below we list online resources for some of the simulation frameworks introduced in this chapter, which provide further information and code downloads:
\begin{itemize}
 \item \eGFRD: \texttt{gfrd.org}
 \item \eCell (incl. \pSpatiocyte, \pGFRD): \texttt{www.e-cell.org}
 \item \ReaDDy: \texttt{www.readdy-project.org} 
 \item \Smoldyn: \texttt{www.smoldyn.org}
 \item \MCell: \texttt{mcell.org} 
 \item {\it MesoRD}: \texttt{mesord.sourceforge.net}
 \item {\it StochSS}: \texttt{www.stochss.org}
\end{itemize}

\section{Summary}
In this chapter we described currently prominent methods used for spatial-stochastic simulation of reaction-diffusion systems,
with a particular emphasis on particle-based simulation schemes, and highlighted their respective advantages and caveats.
We explained Brownian Dynamics as a basic concept for simulating diffusion and reactions of particles,
and pointed out its limitations, especially its intrinsic trade-off between computational efficiency and chemical accuracy.
Within that context, we have introduced detailed balance as an important accuracy criterion for stochastic simulations,
and explained how it can be easily broken by a naive implementation of Brownian Dynamics with reactions.
We demonstrated how accurate Brownian Dynamics algorithms can be derived from the detailed balance criterion,
presenting Reaction Brownian Dynamics and the Reaction Volume Method as illustrating concrete examples.
In the second part of the chapter, we motivated and exemplified the idea of event-driven particle-based stochastic simulation,
focusing on Green's Function Reaction Dynamics (eGFRD), an advanced algorithm that is both exact and highly efficient.
We explained how eGFRD makes use of exact mathematical solutions (Green's functions) of the reaction-diffusion problem 
on simple geometric domains to jump between the relevant simulation events, avoiding the inefficient sampling of particle random walks in between;
we sketched how the Green's functions can be computed analytically and how exact next-event times can be sampled from them.
At the end of the chapter, we presented a brief outlook on current developments 
in the field of stochastic simulation of reaction-diffusion systems, 
in particular mentioning approaches aiming at parallelizing existing schemes,
and efforts to unify particle-based and coarse-grained approaches into hybrid algorithms 
autonomously choosing the best-suited simulation technique depending on local particle density.


\newpage
\section*{Exercises}
\addcontentsline{toc}{section}{Exercises}
\begin{enumerate}
\item Given an interparticle diffusion constant $D=D_1+D_2$ and particle contact radius $R=R_1+R_2$, 
       determine the normalization factor $I_D(\Delta t)$ of the RBD scheme of Sec.~\ref{Sec-Sokolowski-RBD} for an arbitrary time step $\Delta t$
       by explicitly computing the ``overlap integral'' $\int_{|\myvec{r}|\geq R} P^{\rm gen}_{u\rightarrow\Overlap}(\myvec{r},\Delta t) d\myvec{r}$ analytically.
 
 \item Implement a ``naive'' Brownian dynamics scheme in which two spherical particles of species A with identical radius $R_0$ and diffusion constant $D_0$ diffuse in a finite box with volume $L^3$ and periodic boundary conditions. When particles end up overlapping, allow them to react in order to form an immobile particle of species B with radius $2R_0$ at an intrinsic rate $k_a$. Let them also dissociate with a rate $k_d$, always placing the two educt particles at contact and random angle (see Table~\ref{tabExPars} for a suggested set of parameters).
       \begin{enumerate}
        \item Determine a suitable simulation time-step $\Delta t$ from $D_0$ and the reaction rates $k_a$ and $k_d$.
        \item Simulate a sufficiently long trajectory and determine the probability $p_B$ that the two particles are bound, 
              i.e. the fraction of time that the particles spend in the bound state B. 
              Determine $p_B$ also analytically, using mean-field chemical reaction kinetics. 
              Does the analytical result agree with the simulation result?
        \item Repeat the simulations and measurements of $p_B$ with varying time steps, e.g. $3\Delta t$, $2\Delta t$ and $\Delta t/2$. What do you see?
        \item In an improved simulation implementing the Reaction Volume Method, accept the reaction upon particle overlap with the acceptance rate $P^{\rm acc}_{\mathcal{V} \rightarrow b}$
	      derived in Sec.~\ref{Sec-Sokolowski-RBD}, Eq.~(\ref{Eq-Sokolowski-RVMAccRate}), while placing back the dissociated particles uniformly within the reaction volume $\mathcal{V}$.
	      Make sure to choose the reaction volume and time step such that the diffusive displacements of the particle on average are 
	      significantly ($\gtrsim$ 10 times) lower than the width of the reactive shell.
	      Repeat once more the simulations with varying time step. What do you see now?
	\item Use the solution of Exercise 1 to also implement the Reaction Brownian Dynamics scheme, in which the reactions upon particle contact are accepted with rate 
	      $P^{\rm acc}_{\mathcal{O} \rightarrow b}$ [Eq.~(\ref{Eq-Sokolowski-RBDAccRate})], while dissociating particles are placed back at a distance $r$ drawn from the normalized distribution $P_{\Overlap\rightarrow r}(r,\Delta t)r^2/I_D(\Delta t)$. Repeat the simulations with varying time steps once again.
       \end{enumerate}       

 \item Perform the coordinate transform and carry out the separation ansatz for the two-particle case, i.e. starting from the \person{Smoluchowski} equation, Eq.~(\ref{Eq-Sokolowski-Pair}), obtain two separated diffusion equations for the weighted center-of-mass $\myvec{R}$ and the interparticle vector $\myvec{r}$, as in Eq.~(\ref{Eq-Sokolowski-PairSeparated}).
 
 \item Compute analytically the Green's function $p(\myvec{r},t|\myvec{r}_0)$ used in \textit{Single} domains in \eGFRD, i.e. solve the boundary value problem
       for a diffusing particle with radius $\rho$, initially located at the center $\myvec{r}_0$ of a spherical volume of radius $a$ with an absorbing outer boundary, defined by:
       \begin{align}
        \ddt p(\myvec{r},t|\myvec{r}_0) &= D\nabla^2_{\myvec{r}} p(\myvec{r},t|\myvec{r}_0) + \delta(\myvec{r}-\myvec{r}_0)\delta(t-t_0)	\\
        p(\myvec{r},t|\myvec{r}_0) &= 0 \quad\quad\text{for}\quad\quad |\myvec{r}-\myvec{r}_0|=a-\rho
       \end{align}
       In the following we give guidance to first compute the solution for an arbitrary $\myvec{r}_0$,
       taking $\myvec{r}_0\rightarrow 0$ in a second step in order to adapt it to the given problem.
       It is advisable to initially write the solution as
       \begin{align}
	  p(\myvec{r},t|\myvec{r}_0) = p_{\rm free}(\myvec{r},t|\myvec{r}_0) + p_{\rm corr}(\myvec{r},t|\myvec{r}_0)
       \end{align}
       where $p_{\rm free}(\myvec{r},t|\myvec{r}_0)$ is the Green's function for free, unbounded diffusion,
       and $p_{\rm corr}(\myvec{r},t|\myvec{r}_0)$ a solution to the diffusion equation that vanishes for $t=t_0$,
       which must be adapted in a way that $p(\myvec{r},t|\myvec{r}_0)$ fulfills the boundary condition for $t>t_0$.       
       \begin{enumerate}
        \item Rewrite the diffusion equation in a suitable set of spherical coordinates.
	      Convince yourself that the solution only depends on the radial coordinate,
	      and isolate the radial part of the equation, relevant to the given problem.
        \item By Laplace-transforming the radial diffusion equation, show that $\hat p_{\rm corr}(r,s|r_0)$ in Laplace space must have the form $\frac{C}{r} \sinh\left(r\sqrt{s}\right)$,
	      where $C$ is yet undetermined, and $\hat p_{\rm corr}(r,s|r_0)=\int p_{\rm corr}(r,t|r_0) e^{-st} dt$.
	\item Determine $C$ by applying the boundary condition to $\hat p(r,s|r_0)$ in Laplace space.
	      Make use of the Laplace-form of the free solution:
	      \begin{align}
	       \hat p_{\rm free}(r,s|r_0) &= \frac{1}{8\pi rr_0 \sqrt{D s}}\left( e^{-\sqrt{\frac{s}{D}}|r-r_0|} + e^{-\sqrt{\frac{s}{D}}(r+r_0)} \right)
	      \end{align}	      
	\item Convert $\hat p(r,s|r_0)$ back to the time domain by using the inversion theorem (via the \person{Bromwich}/\person{Fourier-Mellin} integral).
	      Use the residue theorem on a complex extension of the function $\hat{p}$, 
	      substituting $\frac{s}{D} \in \mathbb{R}$ by $z \in \mathbb{C}$;
	      employ the fact that all singularities of lie on the negative real axis,
	      and that for a function of the form $\frac{f(z)}{g(z)}$, where $f(z)$ and $g(z)$ are holomorphic functions,
	      the residue at $z_n$ can be obtained as $\lim_{z\rightarrow z_n} \frac{f(z)}{g'(z)}$.
	      Hint: In the given case, there is an infinite number of (periodic) singularities to be taken into account.
	\item Having obtained $p(\myvec{r},t|\myvec{r}_0)$ in the time domain, 
	      take the limit $r_0 \rightarrow 0$ to obtain the desired symmetric Green's function for a particle starting at the center of an absorbing sphere.
	      Hint: Make use of the fact that $\frac{\sin(c x)}{x} \rightarrow c$ for $x \rightarrow 0$.
	\item Integrate the solution $p(\myvec{r},t|0)$ over its domain of support (the whole sphere) in order to obtain the survival probability $S(t)$,
	      i.e. the probability that the particle has not yet left the spherical domain until time $t$.
       \end{enumerate}
\end{enumerate}

\vfill
\begin{table}[h]
\begin{center}
 \begin{tabular}{|l|c|c|}
  \hline
  Simulation box side length	& $L$ 	&	20~$\micron$	\\  
  Particle radius		& $R_0$ &	0.5~$\micron$	\\
  Reduced simulation box volume	& $V$ & = $L^3 - \frac{4}{3}\pi (2R_0)^3$\\
  \hline
  Diffusion constant					& $D_0$ & 5~$\frac{\micron^2}{s}$\\
  Intrinsic association rate upon particle contact	& $k_a$ & 0.01~$\frac{V}{s}$\\
  Dissociation rate					& $k_d$ & 0.01~$\frac{1}{s}$\\
  \hline
 \end{tabular} 
 \caption{Suggested parameters for Exercise 2.}
 \label{tabExPars}
\end{center}
\end{table}


\newpage
\addcontentsline{toc}{section}{References}
\bibliographystyle{qbio}
\bibliography{SOKOLOWSKI}


\end{document}